\documentclass[prb,twocolumn,showpacs,amsmath,amssymb]{revtex4-1}
\usepackage{bm}
\usepackage{dcolumn}% Align table columns on decimal point
\usepackage[dvipdfm]{graphicx}

\def \braket<#1>{\langle{#1}\rangle}
\newcommand  {\img}   {{\rm{i}}}
\newcommand  {\up}    {\uparrow}
\newcommand  {\down}  {\downarrow}

\newcommand  {\dashed}{\protect\mbox{-\! -\! -}}
\newcommand  {\full}{\protect\mbox{-----}}
\newcommand  {\Tr}{\mathop{\mathrm{Tr}}\nolimits}
\newcommand  {\eqn}[1]{(\ref{eqn:#1})}

\renewcommand{\)}     {\right)}
\renewcommand{\_}[1]  {_{\rm #1}}
\renewcommand{\^}[1]  {^{\rm #1}}
\def \v...{\mbox{\scriptsize $\vdots$}}
\def \c...{\mbox{\small $\cdots$}}
\begin{document}

\preprint{preprint}

\title{Conductance distributions in disordered quantum spin-Hall systems}
\author{K. Kobayashi}
 \email{k-koji@sophia.ac.jp}
\author{T. Ohtsuki}
\affiliation{Department of Physics, Sophia University, Tokyo 102-8554, Japan}

\author{H. Obuse}
\affiliation{Department of Physics, Kyoto University, Kyoto 606-8502, Japan}

\author{K. Slevin}
\affiliation{Department of Physics, Osaka University, Osaka 560-0043, Japan}

\date{\today}

%%% abstract %%%
\begin{abstract}
%% 2term conductance in QSHS %%
We study numerically the charge conductance distributions of disordered quantum spin-Hall (QSH) systems using a quantum network model.
 We have found that the conductance distribution at the metal-QSH insulator transition is clearly different from that at the metal-ordinary insulator transition.
 Thus the critical conductance distribution is sensitive not only to the boundary condition but also to the presence of edge states in the adjacent insulating phase.
%% Gpc %%
 We have also calculated the point-contact conductance.
 Even when the two-terminal conductance is approximately quantized, we find large fluctuations in the point-contact conductance.
 Furthermore, we have found a semi-circular relation between the average of the point-contact conductance and its fluctuation.
\end{abstract}
%     AL,       deloc.,   current,  QH transition
\maketitle
%##########################################

%%%%% Introduction %%%%%
\section{Introduction}
%%% QSH %%%
%% IQHE %%
 The integer quantum Hall (IQH) effect
is observed in two-dimensional (2D) electron systems in a strong perpendicular magnetic field.\cite{Klitzing:QHE,Klitzing:review,Huckestein,KOK}
 Both the two-terminal conductance and Hall conductances in the IQH insulating phase are exactly quantized in integer multiples of $e^{2}/h$.
 Transitions between quantized values occur as the Fermi energy passes through the center of a Landau band.
 According to the theory of Anderson localization,
all electrons in 2D systems with broken time-reversal symmetry (unitary universality class) should be localized.\cite{HLN}
 Later, the quantization was understood to be related to the peculiar nature of the conducting states (chiral edge states),
which are characterized by a topologically invariant Chern number.\cite{TKNN}
 The quantum Hall insulator is thus called a topological insulator to be distinguished from an ordinary insulator.

%% QSHE %%
 Recently, the counterpart of the IQH effect in the symplectic symmetry class,
the so-called quantum spin-Hall (QSH) effect,
has been observed in 2D systems with strong spin-orbit interactions in zero magnetic field.\cite{KaneMele:QSHE1,KaneMele:QSHE2,Bernevig:QSHE,Murakami:Bi,Koenig:QSHE}
 The QSH insulator exhibits both quantized charge and spin-Hall conductances
and is a topological insulator that is characterized by the $\mathbb{Z}_2$ topological number.\cite{KaneMele:QSHE2}
%% purpose %%
 The purpose of this paper is to investigate the universal properties (the properties that are independent of microscopic details)
of systems which exhibit the QSH effect.
 To this end, we simulate numerically a suitable quantum network model.\cite{Obuse:QSHnwm,Obuse:QSHnwm2,Avishai}
 We focus on the charge conductance, which is easily accessible in experiments.\cite{Koenig:QSHE}

%% M-I criticality %%
 QSH systems have two kinds of insulating phases:
a QSH insulating phase with edge states,
and an ordinary insulating phase without edge states.
 In addition, the QSH system exhibits a metallic phase sandwiched between the insulating phases.
 By tuning the sample setup,
two Anderson metal-insulator transitions are observed.\cite{Koenig:QSHE}
 Numerical study of the critical exponent that characterizes divergence of the localization length\cite{Onoda07,Obuse:QSHnwm}
and of bulk multifractality of wave functions\cite{Obuse:QSHnwm2}
indicate that the critical properties of QSH systems are the same as those of conventional 2D symplectic systems, 
which is a universality class with time-reversal symmetry and broken spin-rotation symmetry.\cite{HLN}

%% outline for Sec. III %%
 In this work, we investigate the probability distribution functions of the two-terminal charge conductance for disordered QSH systems.
 At Anderson transitions, the distribution functions are sensitive to the universality class and the geometry including boundary conditions\cite{Slevin:BC,Slevin:BClett} of the system,
but are independent of the model and system size.
 We show that the conductance distribution in the metallic and ordinary insulating phases,
and the critical conductance distributions at the metal-ordinary insulator (M-OI) transition of QSH systems 
behave in the same manner as those of the 2D symplectic systems.
 On the other hand, in the QSH insulating phase, the conductance clearly reflects the property of the edge states.
 Remarkably,
the critical conductance distribution at the metal-QSH insulator (M-QSHI) transition is completely different from that at the M-OI transition.
 It is to be noted that the critical conductance distributions depend not only on the symmetry and geometry but also on the presence of edge states in the adjacent insulating phase.
%% Gpc %%
 In addition, we investigate the point-contact conductance, which illustrates the local transport properties of disordered QSH systems.
 Even when the two-terminal conductance is well quantized, 
the point-contact conductance shows the fluctuating nature of the edge states of the QSH insulator.
 This indicates that the conducting state, which is almost free from back scattering, is not always confined to the edge.
 We also demonstrate a peculiar relation between the average and fluctuation of the point-contact conductance.

%%% Outline %%%
 This paper is organized as follows.
%% Model %%
 We first describe the quantum network model for QSH systems in Sec.~\ref{sec:model}.
%% 2-term P(G) %%
 We next study the two-terminal charge conductance in disordered QSH systems in Sec.~\ref{sec:2term}.
%% Gpc %%
 We then study the point-contact conductance in Sec.~\ref{sec:Gpc}.
%% Conclusion %%
 The final section is devoted to the discussions and concluding remarks.

%%%%% Model %%%%%
\section{Model} \label{sec:model}
\begin{figure}[tb]
  \includegraphics[width=70mm]{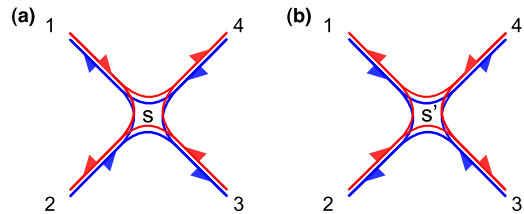}
  \caption{ (Color online)
      Schematics of the (a) $\bm{s}$-type and (b) $\bm{s}'$-type nodes described by the scattering matrices $\bm{s}_i$ and $\bm{s}'_i$, respectively.
      The arrow on upper (under) side of a link indicates the direction of currents
      %The red (blue) arrow on a link indicates the direction of currents
     for up (down) spin.
  }
  \label{fig:node}
 \end{figure}

%%% Model %%%
%% Z2 and CC %%
 To describe the QSH system, we use a $\mathbb{Z}_2$ quantum network model,\cite{Obuse:QSHnwm,Obuse:QSHnwm2,Ryu:Z2nwm}
which is based on the Chalker-Coddington model\cite{CCmodel,KOK} that describes IQH systems.
%% CC model %%
 The Chalker-Coddington model is a regular network of directed current paths.
 It consists of links and nodes,
which correspond to equipotential lines and saddle points of a random potential in the 2D plane, respectively.
%% CC to Z2 %%
 The $\mathbb{Z}_2$ network model can be constructed by introducing the spin degree of freedom into the Chalker-Coddington model so that each link has a Kramers pair,
{\it i.e.}, a pair of current paths for up and down spins propagating in opposite directions.
%%% S-matrix %%%
%% definition %%
 The scattering of the Kramers pairs at an $\bm{s}$-type node at a position $i$ illustrated in Fig.~\ref{fig:node}(a) is described by the $4\times 4$ scattering matrix ${\bm s}_i$,
%--- (eqn: s psi) ---%
   \begin{align} \label{eqn:s_i}
      \begin{pmatrix}
         c\^{out}\_{2\up}\\
         c\^{out}\_{1\down}\\
         c\^{out}\_{4\up}\\
         c\^{out}\_{3\down}
      \end{pmatrix}
       = \bm{s}_i
      \begin{pmatrix}
         c\^{in}\_{1\up}\\
         c\^{in}\_{2\down}\\
         c\^{in}\_{3\up}\\
         c\^{in}\_{4\down}
      \end{pmatrix}, \quad
       \bm{s}_i =
        \Theta_i^{2143}\ {\bm s}\ \Theta_i^{1234},
   \end{align}
%-------------%
where $c_{j\sigma}$ is current amplitude at the link $j$ with the spin $\sigma (=\up,\down)$.
 $\Theta_i^{jklm}$ is defined as
%--- (eqn: Theta) ---%
   \begin{align}
    \Theta_i^{jklm} \equiv
    {\rm diag}\(e^{\img\theta_i^{(j)}},e^{\img\theta_i^{(k)}},e^{\img\theta_i^{(l)}},e^{\img\theta_i^{(m)}}\)
   \end{align}
%-----------------------%
where $\theta_i^{(j)}$ is a random phase associated with the link $j$ for the node $i$,
which is distributed independently and uniformly between $[0,2\pi)$.
 The matrix ${\bm s}$ in Eq. \eqn{s_i} is defined through $2\times 2$ matrices $\bm Q$ and $\bm{I}_2$ as
%--- (eqn: Smat) ---%
   \begin{align} \label{eqn:s_node}
      \bm{s} &=
      \begin{pmatrix}
         \sqrt{p}\, \bm{I}_2 & \sqrt{1-p}\, \bm{Q} \\
         \sqrt{1-p}\, \bm{Q}^T & -\sqrt{p}\, \bm{I}_2
      \end{pmatrix}, \\ \label{eqn:s_QI}
      \bm{Q} &=
      \begin{pmatrix}
         \sqrt{1-q} & \sqrt{q} \\
         -\sqrt{q} & \sqrt{1-q}
      \end{pmatrix}, \ \
      \bm{I}_2 =
      \begin{pmatrix}
         1 & 0 \\
         0 & 1
      \end{pmatrix},
   \end{align}
%-----------------------%
%% Smat parameters %%
where the parameters $p$ and $q$ $(\in [0,1])$ correspond to the chemical potential
and the strength of the spin-orbit interactions that mix up and down spins, respectively.
 The property of $\bm{s}$ and the choice of the phases in Eqs.~\eqn{s_i}~-~\eqn{s_QI} guarantee the time-reversal symmetry.
 The nodes next to the $\bm{s}$-type nodes are the $\bm{s}'$-type [Fig.~\ref{fig:node}(b)].
 The scattering matrix $\bm{s}'_i$ is defined by $90^\circ$ rotation of the matrix $\bm{s}_i$ to make the model isotropic.
%+++++ Fig. node +++++%
%++++++++++++++++++++++%

\begin{figure}[tb]
  \footnotesize
   \begin{tabular}{c@{}ccc}
      $p\to 0$ & $p \simeq 0.5$ & $p\to 1$ \\
      insulator & metal & QSH insulator \\
      \includegraphics[height=29mm]{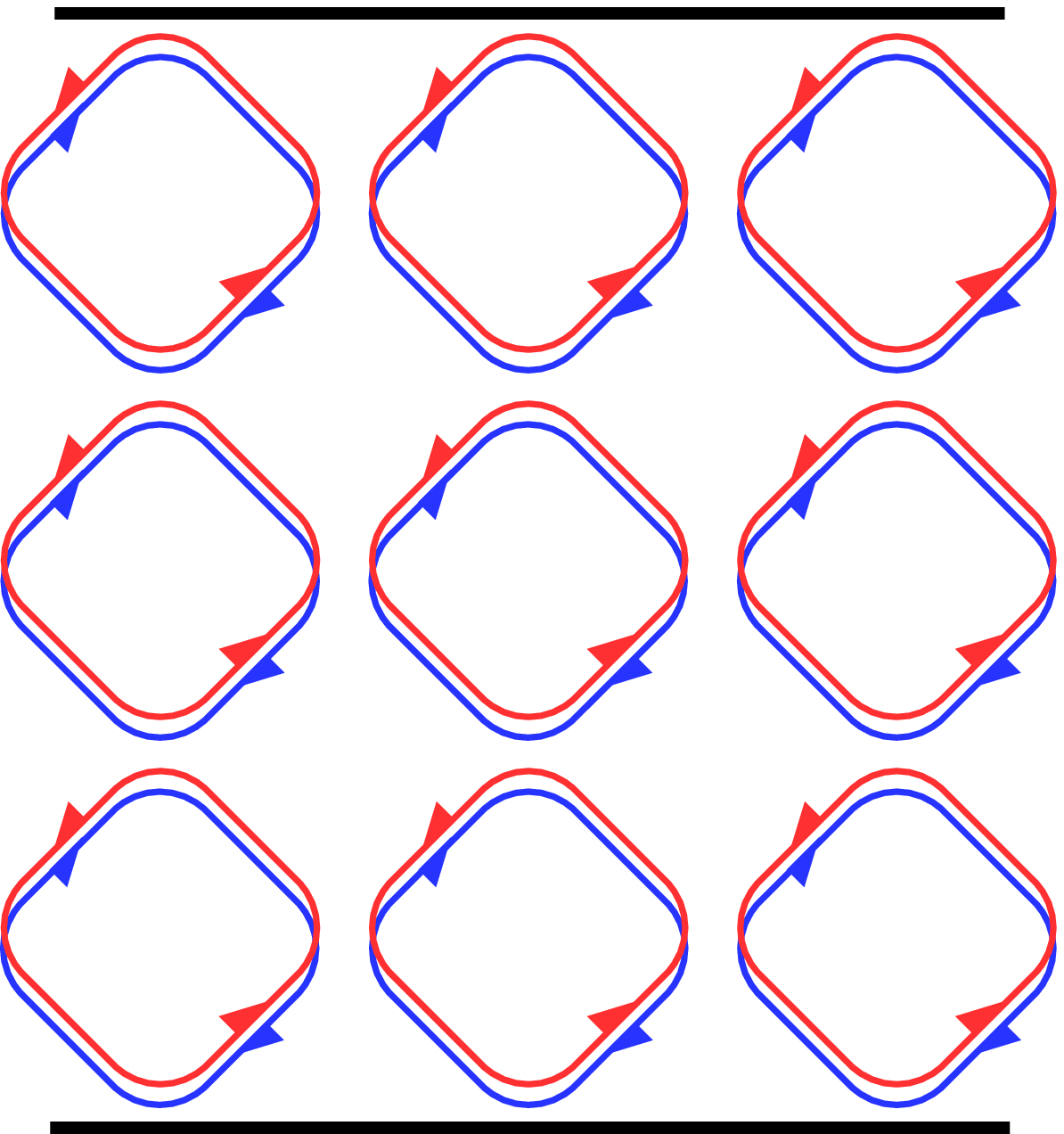}    &
      \includegraphics[height=29mm]{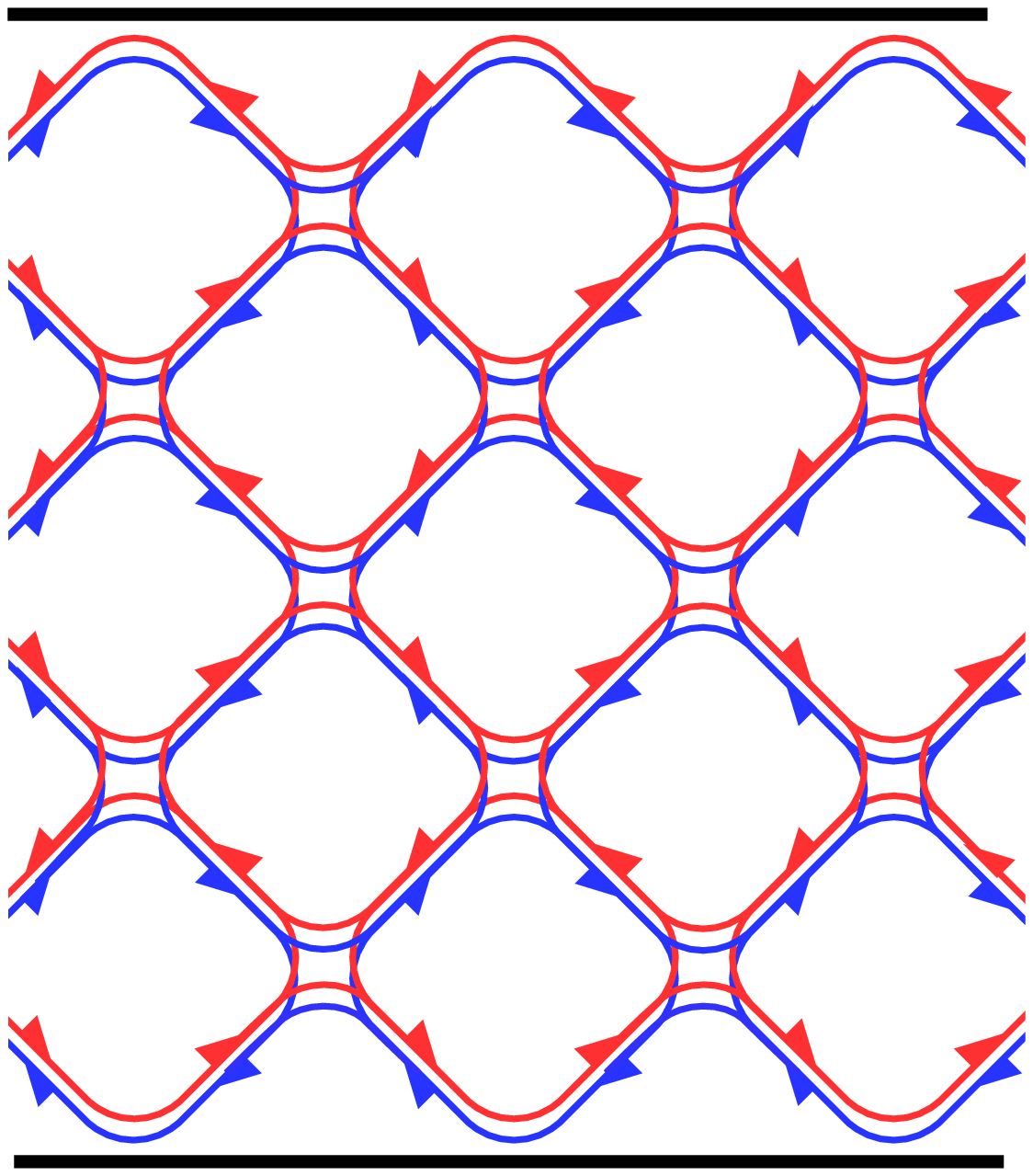} &
      \includegraphics[height=29mm]{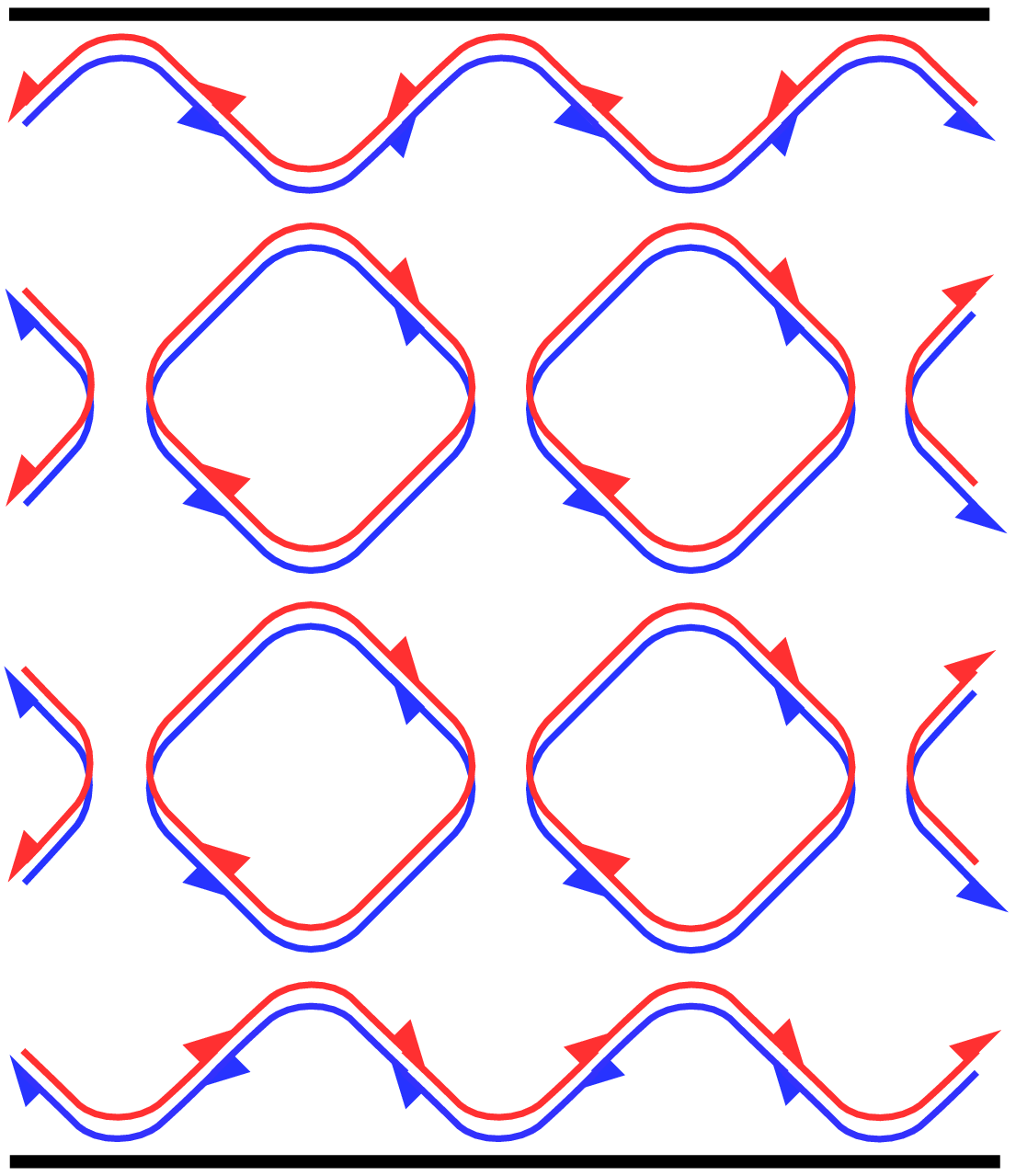}
   \end{tabular}
  \vspace{-2mm}
  \caption{ (Color online)
     Schematics of the $\mathbb{Z}_2$ networks
    in a square geometry with RBC.
     The left and right figures correspond to the network model at $p\to 0$ and $1$, respectively,
    while the middle figure represents the metallic phase at intermediate $p$.
     When $p$ is close to $0$ or $1$, all the bulk states are localized in plaquettes.
     In the QSH insulating phase ($p\to 1$), however,
    two Kramers pairs of conducting channels run along the edges,
    which correspond to the helical edge state.
  }
 \label{fig:QSHNW}
 \end{figure}

%% boundary conditions %%
 Let us explain boundary conditions of the $\mathbb{Z}_2$ network model.
 In this work, we study conductance for the QSH system on which periodic boundary conditions (PBC) or reflecting boundary conditions (RBC) are imposed in the transverse direction.
 (Hereafter, PBC and RBC mean the boundary conditions in the transverse direction to the current
if the direction is not explicitly written.)
 When RBC are imposed,
due to the time-reversal invariance,
a Kramers pair is totally reflected without spin-mixing at a node $i$ located on the boundary,
%--- (eqn: RBC node) ---%
   \begin{align}
    \begin{pmatrix}
       c_{2\up}\^{out} \\
       c_{1\down}\^{out}
    \end{pmatrix}
    =
    e^{\img\theta_i^{(1)}+\img\theta_i^{(2)}} \bm{I}_2
    \begin{pmatrix}
       c_{1\up}\^{in} \\
       c_{2\down}\^{in}
    \end{pmatrix}.
   \end{align}
%-----------------------%
 We assume the arrangement of the nodes as shown in Fig.~\ref{fig:QSHNW}, where
the QSH insulating phase is expected to appear for $p \to 1$,
while the ordinary insulating phase appears for $p \to 0$.

%% system size %%
 The system size of the network is measured in terms of the number of links;
hence, a square system with size $L$ contains $L^2$ links.
 In two-terminal structures for conductance measurements,
$L$ is equivalent to the channel number of Kramers doublets,
which is assumed to be even.
%+++++ Fig. QSHNW +++++%
%++++++++++++++++++++++%

\begin{figure}[tb]
  \includegraphics[width=65mm]{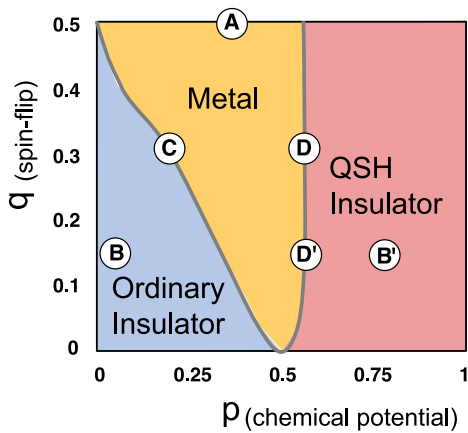}
  \vspace{-2mm}
  \caption{ (Color online)
     Phase diagram (Ref. \onlinecite{Obuse:QSHnwm}) of $\mathbb{Z}_2$ network model with RBC.
     Note that the phase boundaries between metallic and insulating phases
    are the same for networks with PBC.
     For PBC, the ordinary insulating phase replaces the QSH insulating phase.
     Circled letters indicate the points
    considered in the corresponding subsections in Sec.~\ref{sec:2term}.
     The actual values of parameters $(p,q)$ are as follows:
     {\bf A}:  $(0.365,0.500)$,
     {\bf B}:  $(0.039,0.146)$,
     {\bf B'}: $(0.783,0.146)$,
     {\bf C}:  $(0.188,0.309)$,
     {\bf D}:  $(0.561,0.309)$, and
     {\bf D'}: $(0.563,0.146)$.
  }
  \label{fig:phaseDiag}
 \end{figure}

%% Phase Diagram %%
 The detailed phase diagram of the $\mathbb{Z}_2$ network model
has been examined by the finite-size scaling analysis in a quasi-one dimensional geometry\cite{Obuse:QSHnwm}
and is obtained as shown in Fig.~\ref{fig:phaseDiag}.
 Because of its symplectic symmetry,
the system shows a metallic phase sandwiched between two insulating phases.
 When PBC are imposed on the $\mathbb{Z}_2$ network model,
both insulating phases correspond to the ordinary insulating phase.
 On the other hand, when RBC are imposed,
the insulating phase located for larger $p$ exhibits edge states
and becomes the QSH insulating phase as expected.
 Thus, the $\mathbb{Z}_{2}$ network model shows three types of transitions:
the M-OI transition for PBC, the M-OI transition for RBC, and the M-QSHI transition for RBC.
%+++++ Fig. Phase diagram +++++%
%++++++++++++++++++++++%

%%%%%%% 2-term %%%%%%%
\section{Two-Terminal Conductance} \label{sec:2term}
%% settings %%
 In this section, we consider two-terminal charge conductance
for the various phases in the QSH system with a square geometry ($L$ links in both the $x$- and $y$-directions).
%% Method (transfer matrix) %%
 When the ideal leads are attached to the left and right terminals,
the charge transport is described by a $2L\times 2L$ scattering matrix $\bm S$,
%--- (eqn: total S-matrix) ---%
   \begin{align}
      \begin{pmatrix}
         \psi\^{out}\_{L}  \\
         \psi\^{out}\_{R}
      \end{pmatrix}
      =\bm{S}
      \begin{pmatrix}
         \psi\^{in}\_{L}  \\
         \psi\^{in}\_{R}
      \end{pmatrix},
      \quad
      \bm{S}=
      \begin{pmatrix}
         \bm{r} & \bm{t}' \\
         \bm{t} & \bm{r}'
      \end{pmatrix},
   \end{align}
%----------------------%
where $\psi\^{in(out)}\_{L(R)}$ denotes the incoming (outgoing) current amplitude on the left (right) lead,
and $\bm{t}$ and $\bm{t}'$ ($\bm{r}$ and $\bm{r}'$) are $L\times L$ transmission (reflection) matrices.
 By employing the Landauer formula,\cite{KOK} conductance $G$ in units of $e^2/h$ is given by
%--- (eqn: G= Tr tt) ---%
   \begin{align} \label{eqn:Gtt}
      G = {\rm Tr}(\bm t^{\dag}\bm t).
   \end{align}
%-----------------------%
 We have obtained the $\bm t$ matrix by using the transfer matrix method,\cite{KOK}
in which the transfer matrices can be constructed from the scattering matrices $\bm{s}_i$ and $\bm{s}'_i$ in Eqs.~\eqn{s_i}~-~\eqn{s_node} along with boundary conditions.

%%%%% M %%%%%
\subsection{Metallic phase}
%% symplectic metal %%
 First, we focus on the metallic phase.
 It is known that in the metallic phase of systems with symplectic symmetry,
conductance distributions $P(G)$ follow a Gaussian function with constant variance.\cite{LSF:UCF}
 In addition, the averaged conductance $\braket<G>$ increases with logarithm of system size with a universal coefficient as a consequence of the
anti-localization,\cite{HLN}
%--- (eqn: lnL) ---%
   \begin{align} \label{eqn:lnL}
      \braket<G> = G_0 + \pi^{-1}\ln L/l,
   \end{align}
%------------------%
where $G_0$ ($\gg 1$) is the Boltzmann conductance and $l$ denotes the mean free path.

\begin{figure}[tb]
    \includegraphics[width=70mm]{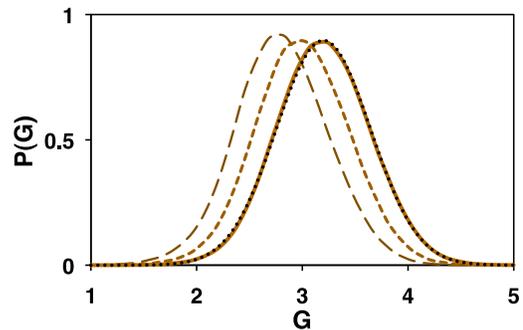}
  \vspace{-2mm}
  \caption{ (Color online)
     Conductance distributions $P(G)$
    in the metallic phase {\bf A} with RBC,
    for $L=64$, $128$, and $256$ (from left to right).
     Distributions keep Gaussian shapes (dotted line) for increasing system size.
  }
  \label{fig:Metal_PG}
 \end{figure}

\begin{figure}[tb]
    \includegraphics[width=70mm]{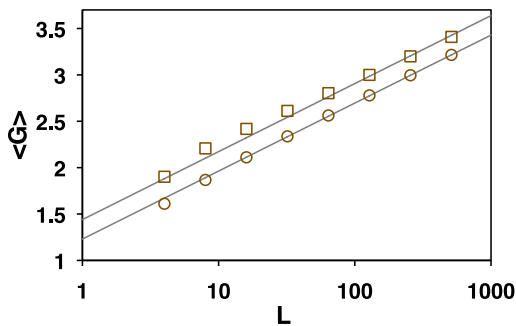}
  \vspace{-2mm}
  \caption{ (Color online)
     Averaged conductances as a function of system size
    in the metallic phase {\bf A} with PBC ($\circ$) and RBC ({\tiny$\square$}).
     The solid lines represent the anti-localization formula Eq.~\eqn{lnL}.
  }
  \label{fig:Metal_G-L}
 \end{figure}

%% calculated P(G) %%
 We have calculated the conductance in the metallic phase of the $\mathbb{Z}_2$ network model at the fixed parameters $p$ and $q$ corresponding to {\bf A} in the phase diagram (Fig.~\ref{fig:phaseDiag}).
 We have confirmed that the distributions are well fitted by Gaussian functions.
 For example, the distributions for systems with RBC are shown in Fig.~\ref{fig:Metal_PG}.
%% HLN %%
 We have also confirmed that the averaged conductance obeys Eq.~\eqn{lnL} irrespective of boundary conditions (Fig.~\ref{fig:Metal_G-L}).
 Therefore, the metallic phase in the QSH systems is qualitatively the same as in the ordinary symplectic systems.
 It is noted that, recently, the anti-localization behavior in the metallic phase of the QSH system is analytically confirmed\cite{Imura} by using a tight-binding model proposed in Ref.~\onlinecite{KaneMele:QSHE1}.
%+++++ Figs. Metal +++++%
%++++++++++++++++++++++%

%%%%% Insu %%%%%
\subsection{Insulating phases}\label{sec:Insu}
%% general %%
 Next, we focus on insulating phases ({\bf B} and {\bf B'} in Fig.~\ref{fig:phaseDiag}).
 It is known, for the ordinary insulating phase, that the averaged conductance exponentially decays as the system size increases and becomes zero in the thermodynamic limit.
 In addition, the distribution function of conductance becomes a log-normal function, i.e., the distribution function of the logarithm of
conductance, $P(\ln G)$, follows a Gaussian function.

\begin{figure}[tb]
    \includegraphics[width=70mm]{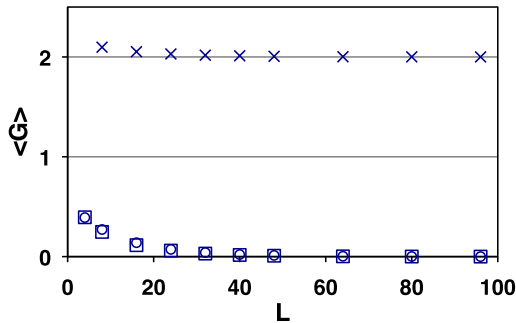}
  \vspace{-2mm}
  \caption{ (Color online)
     Averaged conductance as a function of system size,
    for ordinary insulator {\bf B} with PBC ($\circ$) and RBC ({\tiny$\square$}),
    and QSH insulator {\bf B'} ($\times$).
  }
  \label{fig:Insu_G-L}
 \end{figure}

\begin{figure}[tb]
    \includegraphics[width=70mm]{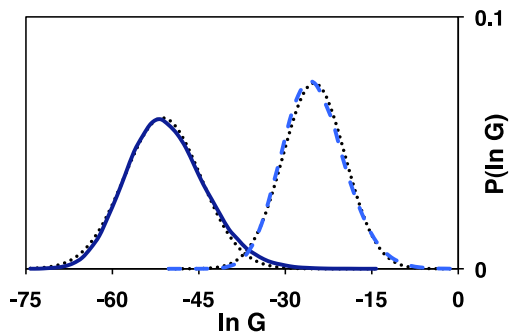}
  \vspace{-2mm}
  \caption{ (Color online)
     Conductance distributions $P(\ln G)$
    in ordinary insulating phase {\bf B} with RBC for $L=256$ (\full) and with PBC for $L=128$ (\dashed).
     For ordinary insulating phase,
    the distribution functions of $\ln G$
    are well fitted by Gaussian functions (dotted line) for any system size.
     For QSH insulating phase, distribution functions $P(G)$ asymptotically approach to the delta function $\delta(G-2)$.
  }
  \label{fig:Insu_PG}
 \end{figure}

%% result %%
 We have calculated the conductance in the ordinary (at {\bf B} in Fig.~\ref{fig:phaseDiag})
and QSH (at {\bf B'}) insulating phases.
 Figure~\ref{fig:Insu_G-L} shows the averaged conductances for each insulating phase rapidly decrease
and converge to different limiting values with increasing system size $L$.
 Indeed, these limiting values of conductance
indicate that the QSH insulating phase has two edge states.\cite{Koenig:QSHE}
 The distribution functions of the logarithm of the conductance $P(\ln G)$
for the ordinary insulator are well fitted by Gaussian functions as shown in Fig.~\ref{fig:Insu_PG}.
 On the other hand, for the QSH insulating phase, $P(G)$ is approximated by the delta function $\delta(G-2)$,
which is an evidence of the edge states.
 The detailed properties of the edge states in the QSH insulating phase are discussed in Sec.~\ref{sec:Gpc}.
%+++++ Figs. Insu +++++%
%++++++++++++++++++++++%

%%%%% M-OI %%%%%
\subsection{Metal-ordinary insulator transition}
\begin{figure}[tb]
  \begin{tabular}{c}
   \includegraphics[width=80mm]{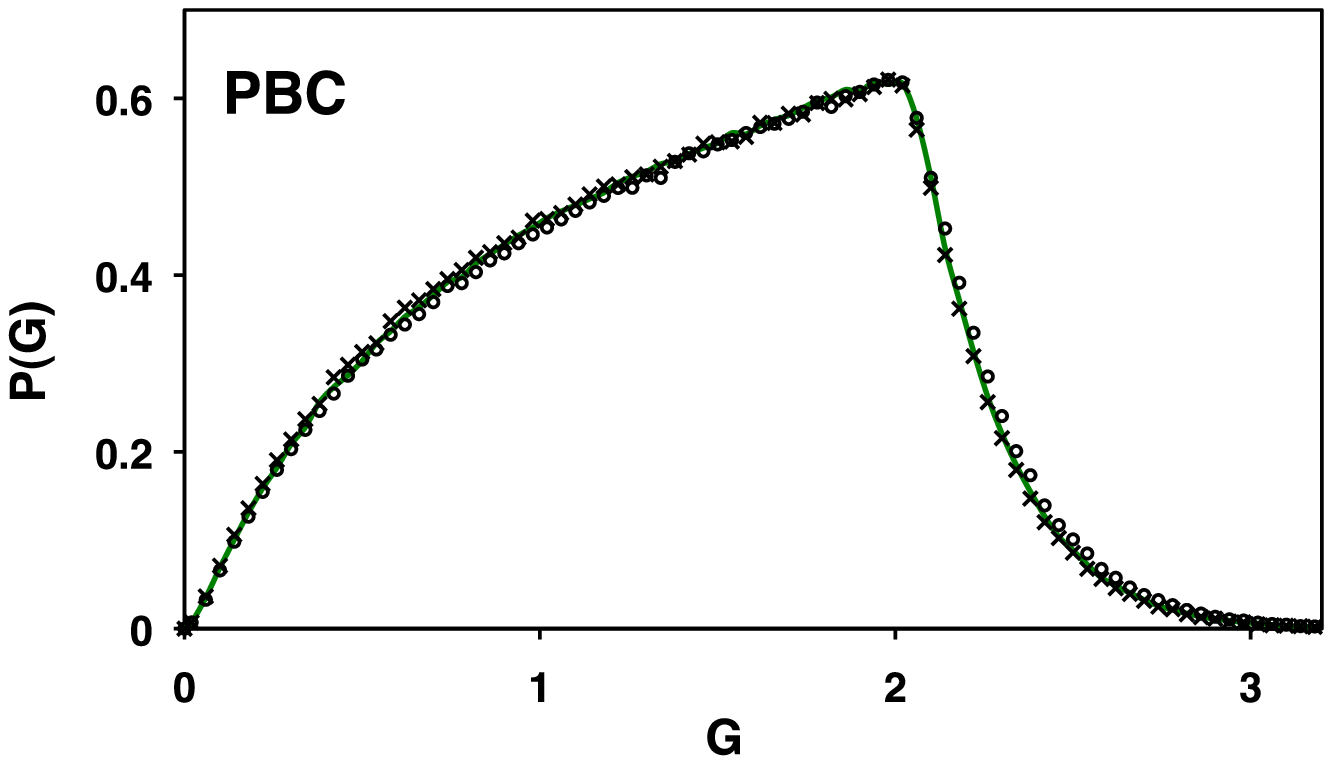} \\
   \includegraphics[width=80mm]{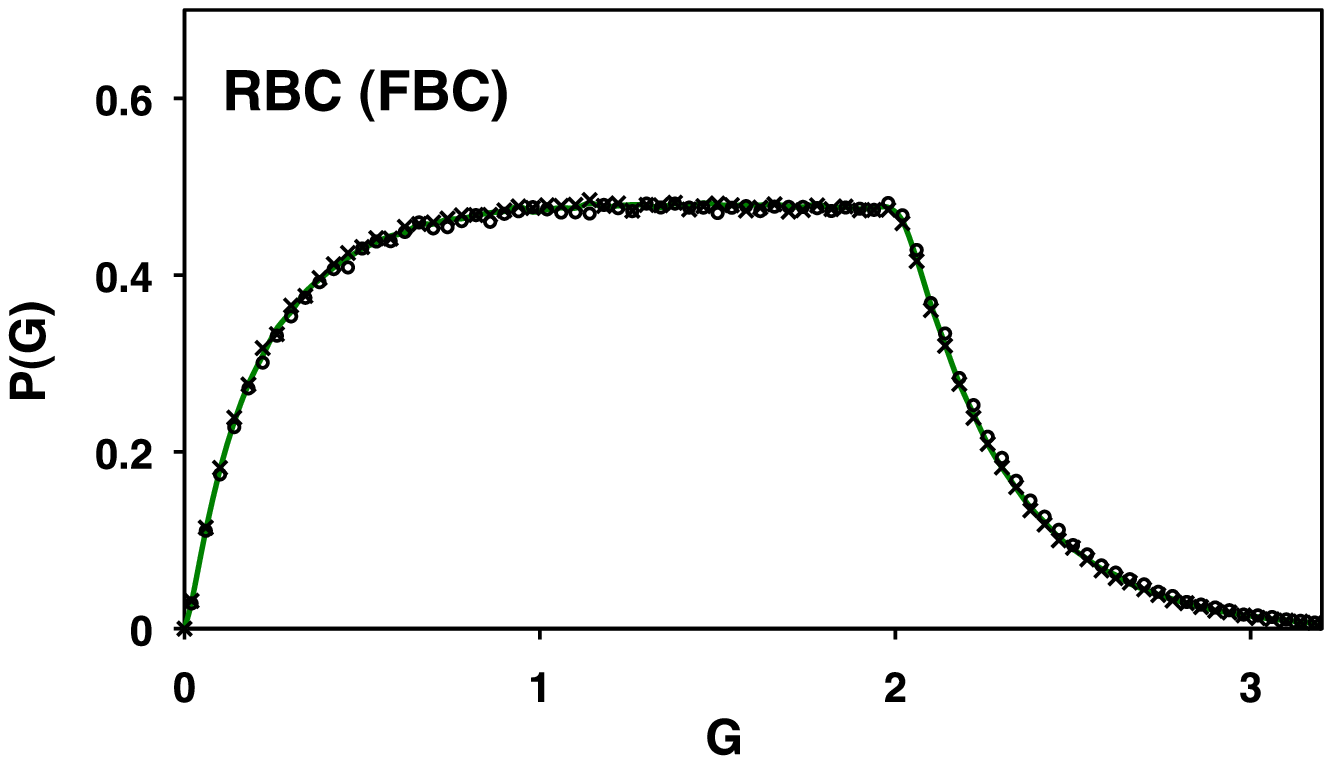}
  \end{tabular}
  \vspace{-2mm}
  \caption{ (Color online)
     Conductance distributions at the M-OI transition point {\bf C}
    calculated on the
    $\mathbb{Z}_2$ network model ($\full$), S2NC model ($\circ$), and
    SU(2) model ($\times$)
   for PBC and RBC [FBC for the SU(2) model].
    More than $10^{6}$ independent random configurations have been realized to draw each curve.
  }
  \label{fig:PG_symplectic}
 \end{figure}

%% universal P(G) for symplectic %%
 Figure~\ref{fig:PG_symplectic} shows the conductance distribution $P(G)$ at the M-OI transition point ({\bf C} in Fig.~\ref{fig:phaseDiag}) in a square geometry with PBC and with RBC.
 For comparison, we also calculated conductance distributions at M-OI transition of the symplectic class by using
a tight-binding model [the SU(2) model\cite{Asada:SU2}] with PBC and fixed boundary conditions (FBC) and
a non-chiral network model (the S2NC model\cite{Merkt}) with PBC and RBC,
which do not show the topologically-nontrivial phase.
 It is found that
the distribution functions at the M-OI transition calculated in the $\mathbb{Z}_2$ network model
coincide reasonably well with those in the other models belonging to the symplectic class.
 We have confirmed that the system size dependence of the distribution functions is negligible for sufficiently large $L$ and
these results are expected to be reproduced everywhere on the line of the same type of transition.
 In fact,
the same distribution has also been obtained at the transition point on the other side of the phase boundary ({\bf D}) for PBC.
 This detail independence of the critical conductance distributions implies that
the distributions are universal and the $\mathbb{Z}_2$ network model, {\it i.e.} the QSH system,
belongs to the same Wigner-Dyson symplectic class as the SU(2) model and the S2NC model.
 This is consistent with the calculation of the critical exponent of the localization length and multifractality\cite{Obuse:QSHnwm,Obuse:QSHnwm2}
%+++++ Fig. P(G) symplectic +++++%
%+++++++++++++++++++++++++++%

%%%%% M-OCI, M-QSHI %%%%%
\subsection{Metal-quantum spin Hall insulator transition}
\begin{figure}[tb]
  \begin{tabular}{c}
   \includegraphics[width=80mm]{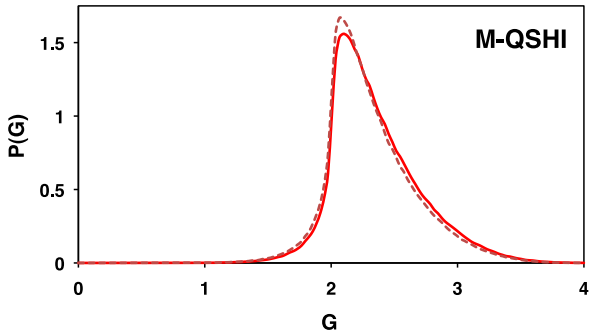}
  \end{tabular}
  \vspace{-2mm}
  \caption{ (Color online)
     Conductance distributions at the M-QSHI transition with $L=256$.
     The distributions at the different points {\bf D} (\full) and {\bf D'} (\dashed) almost coincide with each other.
     $10^6$ samples are calculated for each critical point.
  }
  \label{fig:PG_M-IwithEdge}
 \end{figure}

%% P(G) for M-OCI, M-QSHI %%
 Finally, we focus on the conductance distributions $P(G)$ at the M-QSHI transition points ({\bf D} and {\bf D'} in Fig.~\ref{fig:phaseDiag}).
 As shown in Fig.~\ref{fig:PG_M-IwithEdge},
the conductance distributions at these transitions are completely different from that of the M-OI transition with RBC in Fig.~\ref{fig:PG_symplectic}.
 We emphasize that these distributions are insensitive to details such as system size $L$ ($\gg 1$) or the parameter $(p,q)$ (along the same phase boundary),
 and hence are considered to be universal.
 The distribution at the M-QSHI transition has peak structure around $G=2$,
while they are rather broad at the M-OI transitions (see Fig.~\ref{fig:PG_symplectic}).
 The positions of peaks seem to reflect the helical edge states in insulating phases.
 These results are quantitatively clarified in Table~\ref{tab:critical},
which shows the average $\braket<G>$ and
fluctuation, 
%--- (eqn: sdv) ---%
   \begin{align}
    {\rm sdv}(G)=\sqrt{{\rm var}(G)}=\sqrt{\braket<G^2>-\braket<G>^2}
   \end{align}
%-----------------------%
of the conductance for each transition.
%+++++ Fig M-QSHI, M-OCI +++++%
%+++++++++++++++++++++++++++%
%+-+-+ Table 1 critical +-+-+%
 \begin{table}[tb]
  \caption{
     Averages and fluctuations of the critical conductance in units of $e^2/h$.
  }
  \label{tab:critical}
   \begin{tabular}{r@{}lccc}
    \hline
     \multicolumn{2}{c}{Transition} & $\braket<G>$ & ${\rm sdv}(G)$\\
    \hline
        M-&OI (PBC)  &  $1.42$  &  $0.60$ \\
        M-&OI (RBC)  &  $1.27$  &  $0.65$ \\
        M-&QSHI      &  $2.37$  &  $0.35$ \\
    \hline
   \end{tabular}
 \end{table}
%+-+-+-+-+-+-+-+-+-+-+-+-+-+-%

\begin{figure}[tb]
  \begin{tabular}{cc}
   \includegraphics[width=42mm]{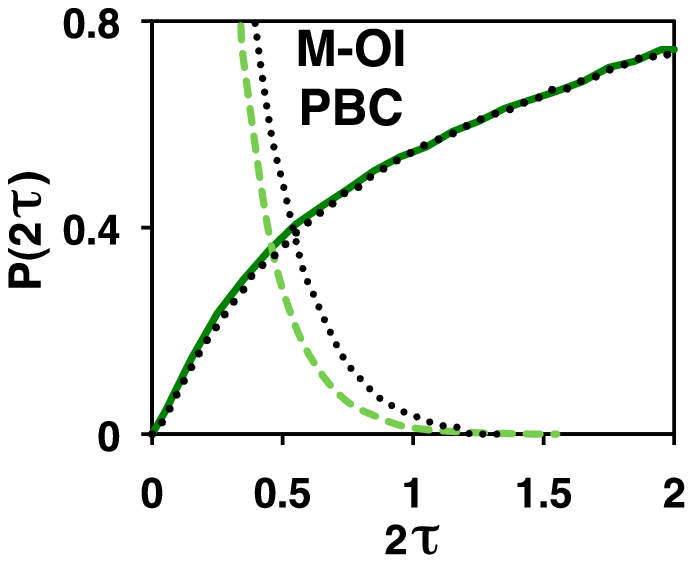} \ &
   \includegraphics[width=42mm]{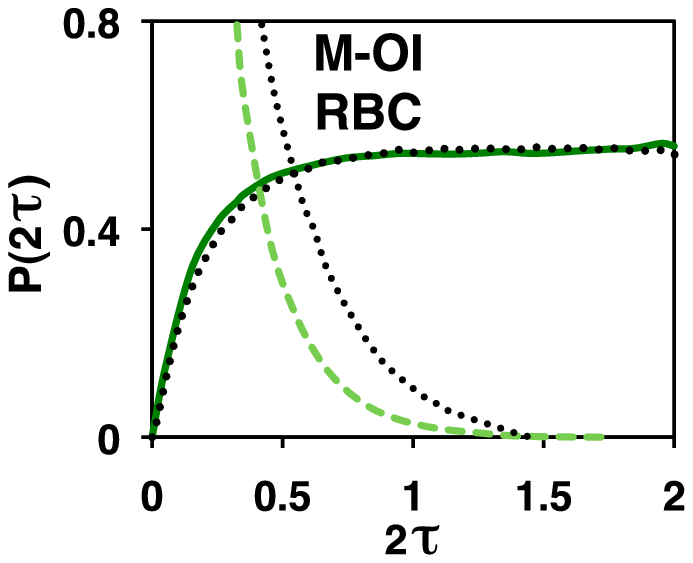}
  \end{tabular}\vspace{-2mm}
  \caption{ (Color online)
     Distribution functions of the largest $P(2\tau_1)$ (\full) and second-largest $P(2\tau_3)$ (\dashed) transmission eigenvalues for M-OI transitions {\bf C} with PBC and RBC.
     Dotted lines are the distribution functions of the conductance $P(G)$ in Fig.~\ref{fig:PG_symplectic} for $G<2$ and 
     for $G>2$, both of which are normalized to be $1$.
  The latter is shifted by $-2$ along the horizontal axis to be compared with $P(2\tau_3)$.     
  }
  \label{fig:Ptau_M-OI}
 \end{figure}

\begin{figure}[tb]
  \begin{tabular}{c}
   \includegraphics[width=52mm]{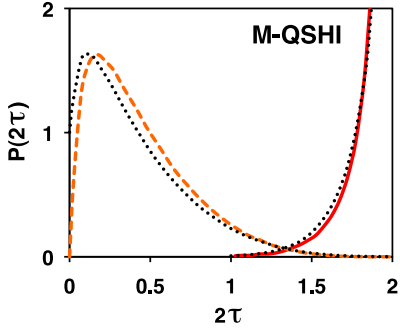}
  \end{tabular}\vspace{-2mm}
  \caption{ (Color online)
     Distribution functions of the largest $P(2\tau_1)$ (\full) and second-largest $P(2\tau_3)$ (\dashed) transmission eigenvalues for M-QSHI transition {\bf D}.
     Dotted lines are the distribution functions of the conductance $P(G)$ in Fig.~\ref{fig:PG_M-IwithEdge} for $G<2$ and
     for $G>2$, both of which are normalized to be $1$.
  The latter is shifted by $-2$ along the horizontal axis to be compared with $P(2\tau_3)$.     
  }
  \label{fig:Ptau_M-QSHI}
 \end{figure}

%%% eigenvalues %%%
 The transmission eigenvalue statistics gives us a better understanding of the results mentioned above.
 The transmission eigenvalue is defined as the set of eigenvalues of $\bm{t}^\dag\bm{t}$ in Eq.~\eqn{Gtt},
$\{\tau_1,\tau_2,\cdots,\tau_L\}$, where we order them according to
%--- (eqn: tau) ---%
   \begin{align}
    \tau_1 \ge \tau_2 \ge \cdots \ge \tau_L.
   \end{align}
%-----------------------%
 We note that all of the eigenvalues are doubly-degenerate.
 Thus,
%--- (eqn: tau_even) ---%
   \begin{align}
    \tau_1 = \tau_2 > \tau_3 = \tau_4 \cdots > \tau_{L-1} = \tau_L.
   \end{align}
%-----------------------%
 Here we focus on the largest and the second largest eigenvalues $\tau_1$ and $\tau_3$.
 We also note that the transmission eigenvalues and the conductance are related as
%--- (eqn: G tau) ---%
   \begin{align}
    G = \sum_{i=1}^{L} \tau_i.
   \end{align}
%-----------------------%
 Figures~\ref{fig:Ptau_M-OI} and \ref{fig:Ptau_M-QSHI} show the probability distribution functions of the largest and the second largest transmission eigenvalues $P(2\tau_1)$ and $P(2\tau_3)$, for each type of transitions.
 Like the IQH transition,\cite{KOK} the largest eigenvalues determine the bare shape of the conductance distributions $P(G)$ for the M-OI transitions with PBC and RBC.\cite{Ohtsuki:eigenSymp}
 The second largest eigenvalues contribute to the tails of $P(G)$ for $G>2$.
%% eigen M-QSHI %%
 For the M-QSHI transition, the doubly-degenerate largest eigenvalues are narrowly distributed near $\tau_1=1$,
which shift $P(G)$ by about two along the $G$-axis,
and the second largest eigenvalues determine the shape of $P(G)$.
 This implies that well conducting channels exist even at the M-QSHI transition,
although there is no edge state.
 These channels, which dramatically change the shape of $P(G)$ at the M-QSHI transition, might be related to the QSH edge states in the adjacent QSH insulating phase.
%+++++ Fig. eigen +++++%
%+++++++++++++++++++++++++++%
%+++++ Fig. eigen +++++%
%+++++++++++++++++++++++++++%

%%%%%%% Point-Contact Conductance %%%%%%%
%%%%% Gpc %%%%%
\section{Point-Contact Conductance} \label{sec:Gpc}

%%% Intro %%%
\subsection{Point-contact conductance}
\begin{figure}[tb]
  \includegraphics[width=70mm]{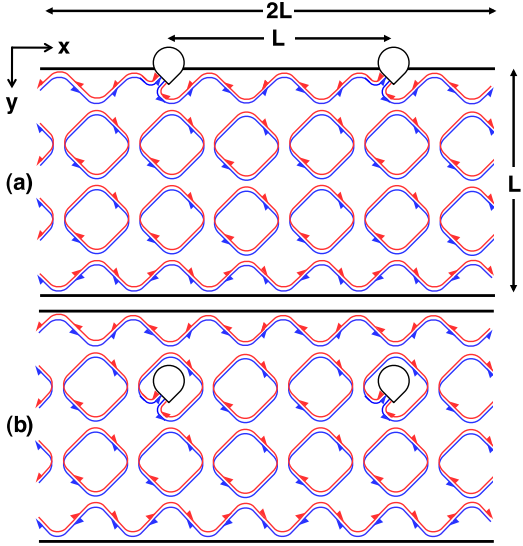}
  \vspace{-4mm}
  \caption{ (Color online)
     Schematics of the $\mathbb{Z}_2$ networks with point contacts
     (a) on the edge, $y\_p=1$, and
     (b) in the bulk region, $y\_p=L/2$.
     Each point contact is connected with a link.
     PBC are imposed on the longitudinal direction.
  }
  \label{fig:pcnet}
 \end{figure}

%% point-contact conductance %%
 To investigate the nature of QSH edge states,
we have calculated the point-contact conductance.
 This is the conductance between two small probes (such as scanning tunneling microscope tips).
%% geometry %%
 For the network model, the point-contact conductance is calculated as the conductance between two links.\cite{Janssen,Klesse,Kobayashi:Gpc}
 We consider a cylindrical geometry ($2L$ links in the $x$-direction and $L$ links in the $y$-direction)
and two point contacts separated by distance $L$ as shown in Fig.~\ref{fig:pcnet}.
%% parameter %%
 The point-contact conductance depends on the position of the contacts in the $y$-direction.
 We assume that both contacts are located at the same distance from one of the edges.
 That is, the contacts are attached at $(x,y)=(0,y\_p)$ and $(L, y\_p)$.

%+++++ Fig. NW Gpc +++++%
%+++++++++++++++++++++++++++%

%%% S-mat calculation %%%
\subsection{Method}
%% Method (S-matrix) %%
 To introduce point contacts into the network,
we cut link $k$ and link $l$ where the contacts are attached.
 We then define incoming current amplitudes
$(c_{k\up}\^{in}, c_{k\down}\^{in}, c_{l\up}\^{in}, c_{l\down}\^{in})$
and outgoing current amplitudes
$(c_{k\up}\^{out}, c_{k\down}\^{out}, c_{l\up}\^{out}, c_{l\down}\^{out})$
on the corresponding links.
 The current amplitudes satisfy the equation
%--- (eqn: evolution ) ---%
   \begin{align} \label{eqn:evolution}
      \begin{pmatrix}
       c_{1\up} \\ c_{1\down} \\[-1mm] \v... \\
       c_{k\up}\^{out} \\ c_{k\down}\^{out} \\[-1mm]
       \v... \\ c_{l\up}\^{out} \\ c_{l\down}\^{out} \\[-1mm] \v... \\[-1mm]
       c_{2L^2\up} \\ c_{2L^2\down}
      \end{pmatrix}
      \!= \tilde{\bm S}\!
      \begin{pmatrix}
       c_{1\up} \\ c_{1\down} \\[-1mm] \v... \\
       c_{k\up}\^{in} \\ c_{k\down}\^{in} \\[-1mm]
       \v... \\ c_{l\up}\^{in} \\ c_{l\down}\^{in} \\[-1mm] \v... \\[-1mm]
       c_{2L^2\up} \\ c_{2L^2\down}
      \end{pmatrix},
   \end{align}
%-------------------------%
where the $4L^2\times 4L^2$ scattering matrix $\tilde{\bm S}$ for all links of a network
consists of the $4\times 4$ scattering matrices $\bm{s}_i$ and $\bm{s}'_i$ in Eqs.~\eqn{s_i}~-~\eqn{s_QI} for a node.
 For given $(c_{k\up}\^{in}, c_{k\down}\^{in}, c_{l\up}\^{in}, c_{l\down}\^{in})$,
the remaining current amplitudes
$(c_{1\up}, \c..., c_{k\up}\^{out}, c_{k\down}\^{out}, \c..., c_{l\up}\^{out}, c_{l\down}\^{out}, \c..., c_{2L^2\down})$
are uniquely determined
by the following set of $4L^2$ simultaneous linear equation with $4L^2$ unknowns
%--- (eqn: lineq) ---%
   \begin{align}
      \begin{pmatrix}
       c_{1\up} \\ c_{1\down} \\[-1mm] \v... \\
       c_{k\up}\^{out} \\ c_{k\down}\^{out} \\[-1mm]
       \v... \\ c_{l\up}\^{out} \\ c_{l\down}\^{out} \\[-1mm] \v... \\[-1mm]
       c_{2L^2\up} \\ c_{2L^2\down}
      \end{pmatrix}
      \!-\tilde{\bm S}\!
      \begin{pmatrix}
       c_{1\up} \\ c_{1\down} \\[-1mm] \v... \\
       0 \\ 0 \\[-0.5mm]
       \v... \\ 0 \\ 0 \\[-0.5mm] \v... \\[-1mm]
       c_{2L^2\up} \\ c_{2L^2\down}
      \end{pmatrix}
      \!= \tilde{\bm S}\!
      \begin{pmatrix}
       0 \\ 0 \\[-1mm] \v... \\
       c_{k\up}\^{in} \\ c_{k\down}\^{in} \\[-1mm]
       \v... \\ c_{l\up}\^{in} \\ c_{l\down}\^{in} \\[-1mm] \v... \\[-1mm]
       0 \\ 0
      \end{pmatrix}.
   \end{align}
%--------------------%
 As a consequence of the structure of these equations,
there is a linear relationship between the incoming and outgoing current amplitudes
%--- (eqn: Spc) ---%
   \begin{align}
      \begin{pmatrix}
       c_{k\up}\^{out} \\
       c_{k\down}\^{out} \\
       c_{l\up}\^{out} \\
       c_{l\down}\^{out}
      \end{pmatrix}
      =
      \begin{pmatrix}
       \bm{r}\_{pc} & \bm{t}'\_{pc} \\
       \bm{t}\_{pc} & \bm{r}'\_{pc}
      \end{pmatrix}
      \begin{pmatrix}
       c_{k\up}\^{in} \\
       c_{k\down}\^{in} \\
       c_{l\up}\^{in} \\
       c_{l\down}\^{in}
      \end{pmatrix}.
   \end{align}
%------------------%
 The point-contact conductance $G\_{pc}$ is given by
%%--- (eqn: Gpc) ---%%
   \begin{align}
      G\_{pc} &= \Tr{({\bm t}^{\dag}\_{pc}{\bm t}\_{pc})},
   \end{align}
%------------------%
in units of $e^2/h$.

%%% Fluctuation of edge state %%%
\subsection{Conductance near edge states}
\begin{figure}[tb]
   \includegraphics[width=75mm]{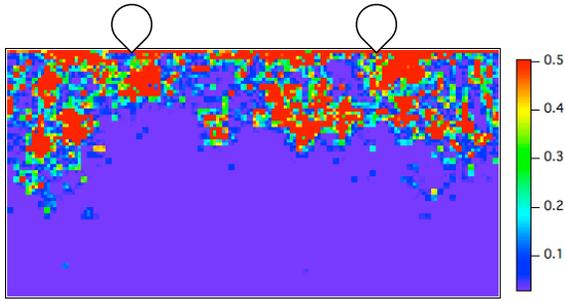}
   \vspace{-2mm}
  \caption{ (Color online)
     Current amplitude $(|c_{i\up}|^2+|c_{i\down}|^2)$ of a $\mathbb{Z}_2$ network
    with $(p,q) =(0.750,0.146)$, for $L=80$.
     The contacts are attached at the upper edge.
  }
  \label{fig:GpcSnap}
 \end{figure}

%% fluctuation in edge state %%
 First we focus on the local property of the edge state in the QSH insulating phase.
 Figure~\ref{fig:GpcSnap} shows the spatial distribution of current for a $\mathbb{Z}_2$ network in the QSH insulating phase.
 The current spreads on the order of $40$ links from the upper edge.
 This is consistent with the estimate of the quasi-1D localization length $(= 19.1\pm 0.2)$ with width $160$, PBC, and $(p,q) =(0.750,0.146)$.
 In this case, the point-contact conductance is not quantized and shows large fluctuations.
 Nevertheless, the two-terminal conductance remains well quantized as shown in Sec.~\ref{sec:Insu}.
 This implies a significant suppression of back scattering, 
once the system width exceeds the localization length.

\begin{figure}[tb]
   \includegraphics[width=70mm]{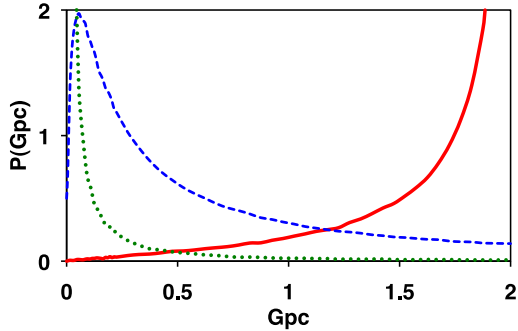}
   \vspace{-2mm}
  \caption{ (Color online)
     Distribution functions of the point-contact conductance
    at the upper edge, $y\_p=1$ (\full), 
    near the edge, $y\_p=3$ (\dashed), and
    at a distance on the order of the localization length, $y\_p=19$ ($\cdots$), 
    with $(p,q) =(0.750,0.146)$, for $L=80$.
    $10^{5}$ samples are realized.
     Although the two-terminal conductance is quantized for corresponding parameters,
    large fluctuations appear for the point-contact conductance.
  }
  \label{fig:PGpc}
 \end{figure}

 Figure~\ref{fig:PGpc} shows
probability distribution functions of the point-contact conductance $P(G\_{pc})$
at $y\_p=1,3,19$ in the QSH insulating phase.
 The distribution $P(G\_{pc})$ for $y\_p=1$ has a peak at $G\_{pc}=2$ with a rather broad distribution.
 In addition, the distributions for $y\_p=3-19$ have broad tails toward $G\_{pc}=2$.
 This indicates that the width of the conducting path associated with a QSH edge state is on the order of the localization length.
%+++++ Fig. snap +++++%
%+++++++++++++++++++++++++++%

%%% ave-sd %%%
\subsection{Relation between average and fluctuation}
\begin{figure}[tb]
  \begin{tabular}{cc}
   \includegraphics[height=45mm]{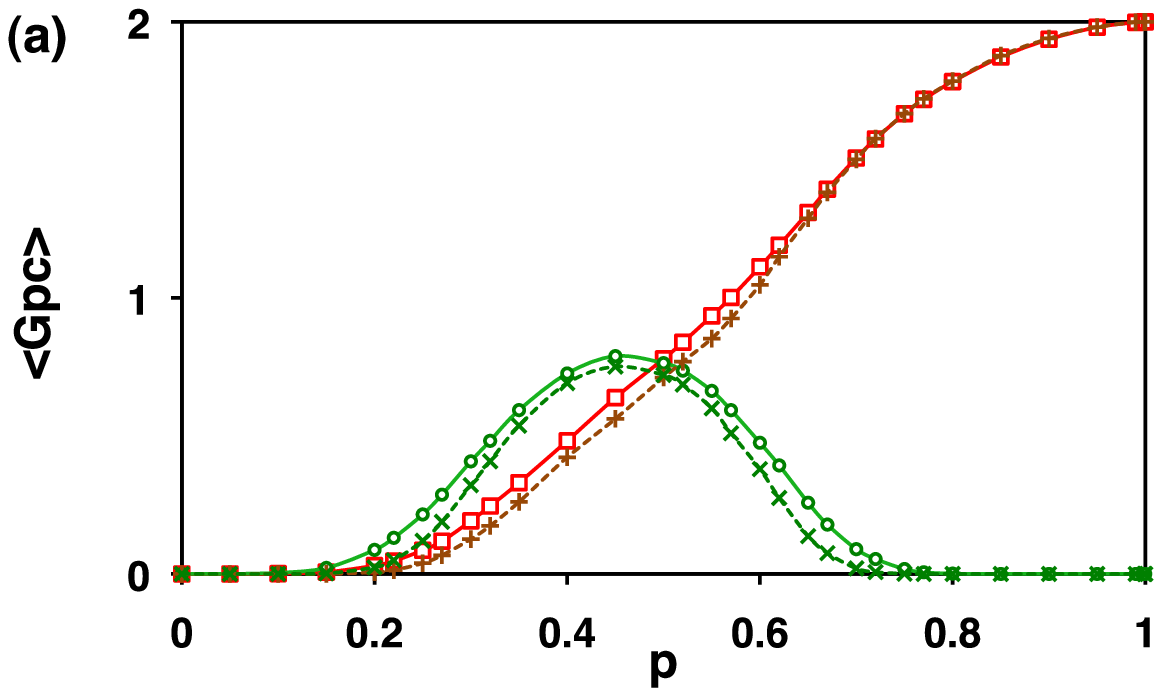} \\
   \includegraphics[height=45mm]{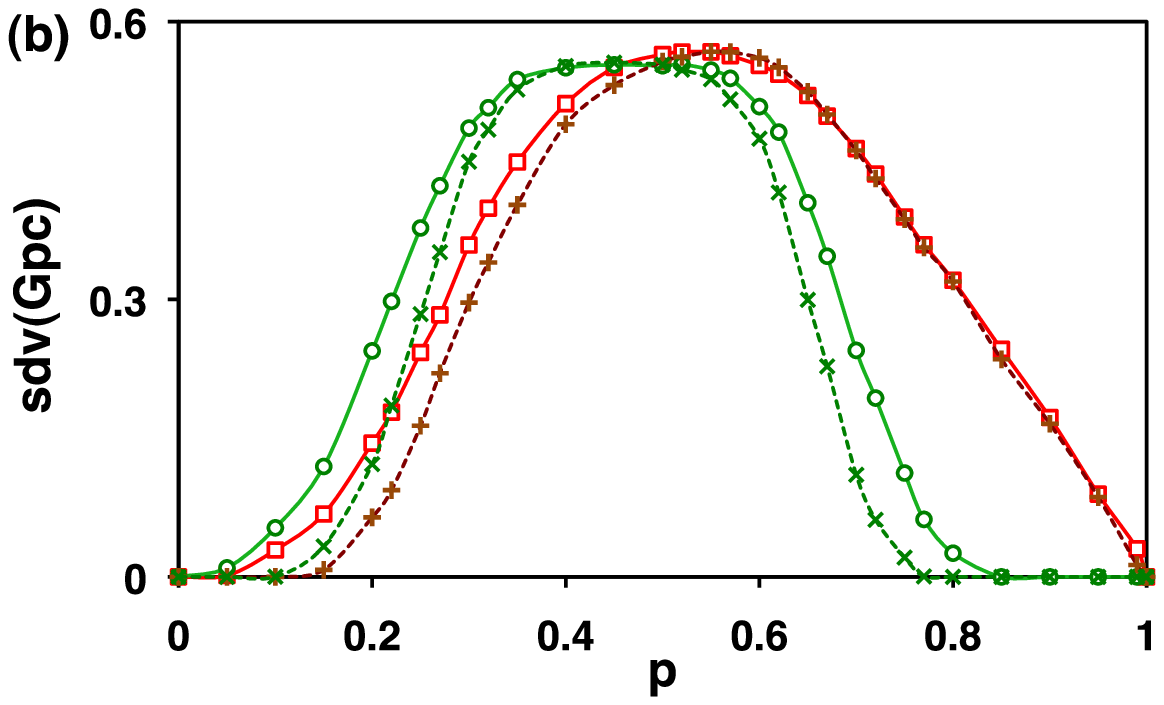}
  \end{tabular}
  \vspace{-2mm}
  \caption{ (Color online)
     (a) Averages and (b) fluctuations of edge ($\square\!:\!L\!=\!80$, $+\!:\!L\!=\!160$) and bulk ($\circ\!:\!L\!=\!80$, $\times\!:\!L\!=\!160$) point-contact conductances with $q=0.146$ as functions of $p$.
     The lines are guide to the eyes.
     The plateau of fluctuation of the bulk conductance corresponds to the metallic phase.
  }
  \label{fig:GpcAve}
 \end{figure}

%% Gpc ave sd %%
 We then see the point-contact conductance as a function of $p$.
 Figure~\ref{fig:GpcAve} shows the average and fluctuation of the edge ($y\_p=1$) and bulk ($y\_p=L/2$) point-contact conductance for $q=0.146$.
 When the bulk conductance becomes small,
two-terminal conductance is well quantized,
since the small bulk conductance means the small inter edge state coupling, which causes the back scattering.
%+++++ Fig. Gpc ave sd +++++%
%+++++++++++++++++++++++++++%

\begin{figure}[tb]
  \includegraphics[width=78mm]{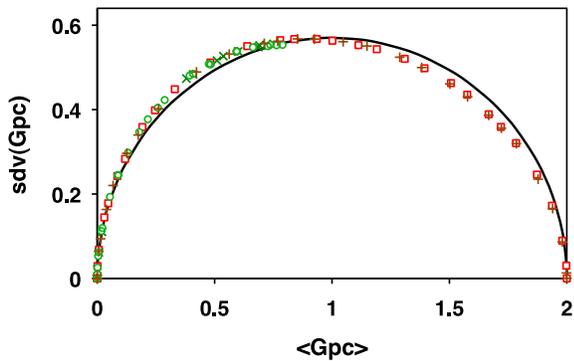}
  \vspace{-2mm}
  \caption{ (Color online)
    Fluctuations as a function of the averaged point-contact conductance
   for edge ($\square\!:\!L\!=\!80$, $+\!:\!L\!=\!160$), $y\_p\!=\!1$, and bulk ($\circ\!:\!L\!=\!80$, $\times\!:\!L\!=\!160$), $y\_p\!=\!L/2$,
   with $q=0.146$.
    This shows a specific relation similar to a semi-circle (\full) between the average and fluctuations
    We have confirmed that
   the same relation is satisfied for different geometry including boundary conditions,
   other contact point ($y\_p=3$), or
   different spin-flip rate ($q=0.309$).
   Statistical errors are smaller than symbols.
   }
  \label{fig:semicirc}
 \end{figure}

%%% Semi-Circle %%%
%\subsection{Semi-circular relation for average-fluctuation}
%% semi-circle %%
 Unlike the two-terminal conductance,
the point-contact conductance shows the system size dependence even at criticality and, then,
it is difficult to find any universal property in the point-contact conductance.
 However, a universal rule appears when we plot the fluctuations as a function of the average of the point-contact conductance for various $p$ (Fig.~\ref{fig:semicirc});
the data points fall on a single curve.
 The curve is approximated by a semi-circle,
although there are small intrinsic deviations.
 The relation is independent of the details of the model and the geometry ($p$, $q$, $L$, or $y\_p$).
 Indeed, we have also confirmed this semi-circler relation for the systems with RBC in both the $x$- and $y$-directions.
%+++++ Fig. semi-circ +++++%
%+++++++++++++++++++++++++++%

%%%%% Discussion %%%%%
\section{Discussion and Concluding Remarks}
%%% 2term %%%
%% Conclusion %%
 In this paper, we have studied the transport properties of QSH systems by numerical calculations on the $\mathbb{Z}_2$ network model.
 We have shown that the conductance distributions at transition points are sensitive to the type of transition, {\it i.e.}, the presence or absence of edge states in the adjacent insulating phase,
and found the universal conductance distribution for the M-QSHI transition.
 We have also shown that the universal conductance distributions for conventional symplectic systems with PBC and RBC
are reproduced in the M-OI transition of the $\mathbb{Z}_2$ network model.
 This is consistent with Ref.~\onlinecite{Obuse:QSHnwm2}, where the boundary multifractality has been shown to be sensitive to the existence of the edge states.

\begin{figure}[tb]
  \includegraphics[width=75mm]{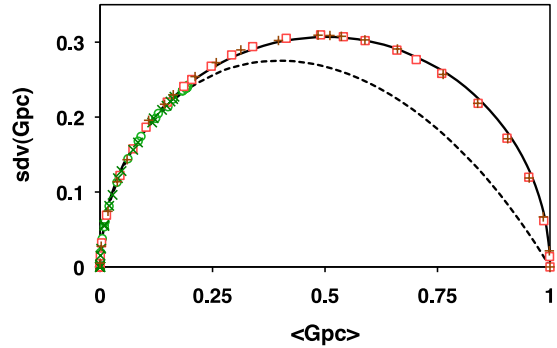}
  \vspace{-2mm}
  \caption{ (Color online)
    Fluctuations as a function of the averaged point-contact conductance
   in the Chalker-Coddington model (IQH system).
    Plots for both edge ($\square\!:\!L\!=\!80$, $+\!:\!L\!=\!160$), $y\_p\!=\!1$, and bulk ($\circ\!:\!L\!=\!80$, $\times\!:\!L\!=\!160$), $y\_p\!=\!L/2$,
   agree with the semi-circular relation (\full).
    The result for the single channel DMPK equation (Refs. \onlinecite{Gertsenshtein} and \onlinecite{Papanicolaou}) (\dashed) agrees with the semi-circle only for small averaged conductance.
  }
  \label{fig:QHsemicirc}
 \end{figure}

%%% Gpc %%%
%% edge fluctuation %%
 In the QSH insulating phase, the point-contact conductance fluctuates remarkably near the edge,
while the two-terminal conductance is well quantized.
%% semi-circle %%
 We have found a universal relation between the average and fluctuation of the point-contact conductance.
%% QHS %%
 The similar relation was suggested for the two-terminal conductance in IQH systems.\cite{Ando:QHf}
 Motivated by this, we have confirmed the relation for the point-contact conductance in IQH systems (Fig.~\ref{fig:QHsemicirc}) and found it is closer to a semi-circle.
 We have also calculated the conductance for diffusive transport in a single-channel quantum wire based on the Dorokhov-Mello-Pereyra-Kumar (DMPK) equation.\cite{Gertsenshtein,Papanicolaou,DMPK:D,DMPK:MPK}
 The curve (Fig. \ref{fig:QHsemicirc}) agrees with our data in the small conductance region,
but deviates from the semi-circle for large conductance, which can be achieved only by the edge state.
 The semi-circular relation, therefore, may be a characteristic of the topological insulators.
 Recently, much experimental progress have been made in developing local probes\cite{Hashimoto:STS,Otsuka:QD} for IQH systems.
 We expect that this relation is experimentally accessible since this relation is independent of the parameters and the boundary conditions.
%+++++ Fig. QHsemicirc +++++%
%+++++++++++++++++++++++++++%

 The above conclusions are mostly based on the network model, the details of which are different from realistic samples.
 However, since the properties calculated here are expected to be universal (see for example, Fig. \ref{fig:PG_symplectic}, where excellent agreement of the tight-binding model with the network model is demonstrated),
we expect our results can be verified by experiments.

%%%%%%%%%%%%%%%%%%%%%%%%%%%%%%%%%%%%%%%%

\begin{acknowledgments}
 We thank M. Yamamoto for fruitful discussions.
 We thank Y. Kohyama for technical advice.
 This work was supported by KAKENHI No. 18540382.
 H.O. is supported by Grant-in-Aid for JSPS Young Scientists.
\end{acknowledgments}

\end{document}